\documentclass{svproc}
\usepackage{graphicx}
\usepackage{url} 

\usepackage[backend=biber,style=apa]{biblatex}
\usepackage[spanish]{babel}
\usepackage{csquotes}
\usepackage{amsmath}
\usepackage{newunicodechar}
\usepackage{float}
\usepackage{setspace}
\begin{document}
\mainmatter

\title{Gestión de Recursos Cuánticos en la Era NISQ: Implicaciones y Perspectivas desde la Ingeniería de Software}

\titlerunning{Gestión de recursos cuánticos en la era NISQ}

\author{Marcos Guillermo Lammers\inst{1} 
\newline Email: \texttt{marcos.lammers@lifia.info.unlp.edu.ar}
\newline {ORCID: \url{https://orcid.org/0009-0007-3954-2252}}
\newline
\and Federico Hernán Holik\inst{2}
\newline Email: \texttt{holik@fisica.unlp.edu.ar}
\newline {ORCID: \url{https://orcid.org/0000-0002-6776-5281}}
\newline
\and Alejandro Fernández\inst{1}
\newline Email: \texttt{alejandro.fernandez@lifia.info.unlp.edu.ar}
\newline {ORCID: \url{https://orcid.org/0000-0002-7968-6871}}
}

\authorrunning{Lammers M., Holik F., Fernández A.}

\tocauthor{Marcos Guillermo Lammers, Federico Hernán Holik, Alejandro Fernández}

\institute{
\inst{} Universidad Nacional de La Plata, Facultad de Informática, Laboratorio de Investigación y Formación en Informática Avanzada (LIFIA), Argentina \and
\inst{} Universidad Nacional de La Plata, Instituto de Física (IFLP-CCT-CONICET), Argentina
}

\maketitle

% Resumen y palabras clave en español
\begin{abstract}
Las computadoras cuánticas suponen un avance tecnológico radical en la forma de procesar la información al utilizar los principios de la mecánica cuántica para resolver problemas muy complejos que exceden las capacidades de los sistemas clásicos. No obstante, en la actual era NISQ (dispositivos cuánticos de escala intermedia ruidosa) el hardware disponible presenta diversas limitaciones, como un número acotado de qubits, elevadas tasas de error y tiempos de coherencia reducidos. La gestión eficiente de los recursos cuánticos —tanto físicos como lógicos— cobra especial relevancia en el diseño y despliegue de algoritmos cuánticos. En este trabajo analizamos el rol de los recursos en los distintos usos actuales de los dispositivos NISQ, identificando su relevancia e implicaciones para la ingeniería de software cuántico.

\vspace{0.5em}

\textbf{Palabras clave:} Computación Cuántica, Gestión de Recursos Cuánticos, Ingeniería de Software Cuántico, Corrección de Errores, Estimación de Recursos Cuánticos (QRE)
\end{abstract}

% Título en inglés
\begin{center}
\textbf{Quantum Resource Management in the NISQ Era: Implications and Perspectives from Software Engineering}
\end{center}

% Abstract y keywords en inglés
\begin{abstract}
Quantum computers represent a radical technological breakthrough in information processing by leveraging the principles of quantum mechanics to solve highly complex problems beyond the reach of classical systems. However, in the current NISQ era (noisy intermediate-scale quantum devices), the available hardware presents several limitations, such as a limited number of qubits, high error rates, and short coherence times. Efficient management of quantum resources —both physical and logical— is especially relevant in the design and deployment of quantum algorithms. In this paper, we analyze the role of resources in current uses of NISQ devices, identifying their relevance and implications for quantum software engineering. With this contribution, we aim to strengthen the field of Quantum Resource Estimation (QRE) and move toward scalable and reliable quantum software development.

\end{abstract}

\section{Introducción}
\vspace{-0.6em} 
La computación cuántica es un campo de investigación y desarrollo interdisciplinar que promete aplicaciones relevantes a futuro en áreas tales como la ciberseguridad, la resolución de problemas de optimización en la industria y las finanzas, y la simulación de moléculas para la industria química y farmacéutica. 

Ya existen computadoras cuánticas, algunas de las cuales están disponibles comercialmente. En este trabajo, vamos a distinguir entre el concepto de \textit{computadora cuántica universal y tolerante a errores} (CCUTE), y los dispositivos disponibles actualmente, a los cuales nos referiremos como dispositivos ruidosos o imperfectos. En la actualidad, no existen CCUTEs lo suficientemente robustas como para realizar aplicaciones comerciales relevantes tales como aplicar el algoritmo de Shor (\cite{shor1994}) para quebrar una clave criptográfica. Los expertos se refieren a los dispositivos disponibles en la actualidad con el acrónimo ''NISQ'', en referencia al concepto de Noisy Intermediate Scale Quantum (dispositivos cuánticos ruidosos de escala intermedia). Si bien los avances de los últimos años han sido importantes (\cite{sycamore2019, Zuchongzhi2024, ibm2025, iqm2030}), la mayoría de los expertos coincide en que estamos a varios años de la era de las CCUTEs y de las aplicaciones comerciales relevantes. Algunos autores estiman que podríamos estar a décadas de la era de las CCUTEs capaces de resolver problemas comercialmente relevantes.

La era NISQ es un término acuñado por John Preskill (\cite{Preskill2018}). Allí, el autor se refiere a computadoras de entre 50 y 100 qubits. Las describe como computadoras limitadas por el ruido a circuitos cuánticos poco confiables y reducidos. Pueden ser útiles en experimentos e investigaciones relacionadas con la física cuántica. Actualmente existen computadoras accesibles de 156 qubits y otras prometidas de más de 1000 (\cite{ibm2025}). Sin embargo, el ruido sigue siendo un problema difícil de afrontar. Algunas computadoras actualmente disponibles basadas en trampas de iones tienen qubits menos ruidosos y compuertas de mayor fidelidad, pero tienen del orden de 50 qubits físicos. Sumado a esto, la dispersión del hardware existente, la falta de estándares, el elevado costo de los desarrollos y las dificultades técnicas en la escalabilidad, hacen que no sea viable la aplicación de algoritmos y circuitos cuánticos de interés comercial. Existen muchas promesas para computadoras tolerantes a fallos en el futuro (\cite{iqm2030,ibm2025,microsoft2025}). Muchos estados y empresas están trabajando en su desarrollo. Hasta ahora, a pesar de los avances, las computadoras tolerantes a fallos con un número razonablemente grande de qubits no se han hecho realidad. Sin embargo, los prototipos actualmente disponibles encuentran un uso creciente destinado a fines de investigación y desarrollo. Una ingeniería de software que tenga presente sus limitaciones nos permitirá facilitar su uso durante la era NISQ, así como estar mejor preparados para posibles actualizaciones tecnológicas en los próximos años.

Si bien no se observa en la actualidad un uso comercial extendido de los dispositivos NISQ, es importante remarcar que en los últimos años se ha desarrollado progresivamente una comunidad que involucra a distintos tipos de usuarios. Estos se concentran en equipos de investigación de instituciones públicas y empresas. Estas investigaciones son de diversa índole y pueden clasificarse (al menos) en tres grupos. Un primer grupo se enfoca en mejorar los dispositivos existentes, detectando y corrigiendo sus errores e identificando qué pasos hay que seguir en su desarrollo a futuro. Un segundo grupo se refiere al posicionamiento temprano de empresas, estados y grupos académicos respecto de la resolución de problemas a futuro (un ejemplo de esto es el desarrollo de equipos de investigación de empresas farmacéuticas que buscan aplicaciones en el mediano plazo). Finalmente, encontramos un tercer grupo que se concentra en estudiar si es posible obtener algún tipo de ventaja, ya sea comercial o de investigación pura, en el uso de dispositivos NISQ. Naturalmente, la separación entre estos tres grupos es difusa y existe un gran solapamiento entre las distintas tareas. Estos proyectos son evidencia de la existencia de una comunidad significativa de usuarios de dispositivos NISQ para los cuales el desarrollo de herramientas de software para alcanzar sus objetivos de investigación y desarrollo se ha vuelto un desafío cotidiano. Esto implica la necesidad de invertir una cantidad considerable de recursos en el desarrollo de dichas herramientas. Atendiendo a los avances de los últimos años, es importante tener en cuenta que este estado de cosas podría extenderse por varios años e incluso por décadas.

En ese contexto, podemos formular de forma precisa la pregunta que abordaremos en este trabajo: ¿Cuáles son las implicaciones para la ingeniería del software cuántico del hecho de que los usuarios de computadoras cuánticas tengan que lidiar con dispositivos NISQ, los cuales poseen recursos limitados y están sujetos a errores? La relevancia de esta pregunta está vinculada a otro hecho: la mayoría de las investigaciones en el campo conocido como \textit{Ingeniería de Software Cuántico} se enfocan en CCUTEs. Es decir, se concentran en el estudio de dispositivos por venir sin tener en cuenta los desafíos que plantean las investigaciones que llevan a cabo los usuarios de dispositivos NISQ. La propuesta de este trabajo es identificar y sistematizar algunas de las prácticas de desarrollo que son relevantes para la ingeniería del software orientado al uso de dispositivos NISQ.

Este artículo se organiza de la siguiente manera: En la sección \ref{s:Ingeniería_de_software} se discute el rol de los recursos en la ingeniería del software. La sección \ref{secction:rcq} se adentra en el mundo de la computación cuántica, poniendo el eje en el rol de los recursos en los dispositivos NISQ, y los problemas vinculados a la caracterización de su performance y su calibración. La sección \ref{section:related-work} hace un breve repaso de los esfuerzos disponibles en trabajos publicados  orientados a la caracterización y estimación de los recursos disponibles en procesadores cuánticos. Como resultado de ese análisis, se observa que la mayoría de los desarrollos están enfocados en CCUTEs. La sección \ref{s:Recursos_en_NISQ} hace un análisis del rol de los recursos en los distintos usos que se hacen de los dispositivos NISQ en la actualidad, identificando su relevancia e implicancias para la ingeniería del software orientada al uso de computadoras cuánticas. Finalmente, la sección \ref{s:Conclusiones} expone conclusiones y desafíos a abordar en futuros trabajos.
\vspace{-0.6em} 
% Secciones 
\section{El rol de los recursos en la ingeniería de software}\label{s:Ingeniería_de_software}
\vspace{-0.6em} 
La ingeniería de software, como disciplina, tiene entre sus objetivos principales la gestión eficiente de los recursos disponibles, reconociendo que estos son finitos y de ellos depende el éxito de los proyectos. Se ocupa de todas las fases del desarrollo de software, desde su diseño hasta su mantenimiento (\cite{sommerville2016}).

Para la ejecución de cualquier software es necesario tener en cuenta distintos tipos de recursos, tales como el tiempo de procesamiento, la memoria, archivos y dispositivos de entrada/salida. Estos recursos pueden ser requeridos por el software al instante de su creación o durante su ejecución (\cite{silberschatz2006}). Esto significa que, desde los inicios, los ingenieros deben organizar, entre otras cosas, la utilización de los recursos de hardware y software para prevenir atascos y asegurar un desempeño óptimo. Como menciona Sommerville, en la ingeniería de software tradicional, el desafío consiste en crear técnicas para generar software confiable que posea la suficiente adaptabilidad para gestionar la heterogeneidad (\cite[p. 24]{sommerville2016}), lo que incluye la capacidad para adaptarse a las limitaciones del hardware existente.

Los recursos de hardware influyen directamente en decisiones como la selección de arquitecturas de software y la optimización del rendimiento en cualquier tipo de dispositivo. Esto no es diferente en el uso de computadoras cuánticas. El hardware debe utilizarse de manera eficiente para cumplir con los requisitos de rendimiento y confiabilidad. El desafío principal es encontrar el punto de equilibrio entre los recursos escasos y los objetivos del proyecto, asegurando que el software sea eficiente, escalable y capaz de operar dentro de las limitaciones físicas disponibles. Para verificar esto, necesitamos conocer de manera confiable el hardware disponible y las necesidades del software.
\vspace{-0.6em} 
\section{Recursos de las computadoras cuánticas}
\label{secction:rcq}
\vspace{-0.6em} 
A grandes rasgos podemos clasificar a los recursos de una computadora cuántica de acuerdo a sus niveles \textit{físicos} y \textit{lógicos}. Por un lado, tenemos los recursos físicos. Ejemplos de esto son el número de qubits físicos, su tasa de error y el tiempo de coherencia, el nivel de entrelazamiento que sea posible alcanzar, la conectividad de los qubits, y la fidelidad de las compuertas nativas. Por otro lado, están los recursos lógicos, como por ejemplo, el número de qubits lógicos, el conjunto de compuertas disponibles en la plataforma de acceso al dispositivo, la máxima profundidad del circuito cuántico soportada, los algoritmos cuánticos pre-implementados, los protocolos de corrección de errores, limitaciones en el número y naturaleza de mediciones, y las herramientas de estimación de errores, entre otros. Estos recursos tienen que ser tenidos en cuenta en el modelo específico de la computación cuántica. Esa diferencia respecto a la computación clásica hace necesario un enfoque especial de ingeniería de software.

En computación cuántica, un gran número de qubits físicos suele ser considerado como una ventaja en lo que respecta a tener una mayor capacidad de procesamiento de datos. En teoría, una mayor cantidad de qubits físicos permitiría lograr más qubits lógicos en los cuales implementar protocolos de corrección de errores cuánticos. Es importante destacar que, aunque un gran número de qubits físicos es en principio deseable, también hay que tener en cuenta su calidad. Los qubits físicos con alta tasa de error no serán útiles, incluso aunque fueran numerosos. Este hecho refuerza la importancia de la calidad de los qubits como un recurso clave (\cite{Preskill2018}). La tasa de error está íntimamente ligada al tipo de computadora cuántica (QC, por sus siglas en inglés) utilizada. Para lograr ejecutar algoritmos, obtener resultados confiables y que el sistema sea escalable, es necesario alcanzar una baja tasa de error.

El tiempo de coherencia nos dice cuánto tiempo un qubit puede mantener su estado cuántico antes de que la decoherencia lo degrade y se pierda la información. Para que un algoritmo cuántico se ejecute correctamente, el tiempo de coherencia tiene que ser lo suficientemente largo como para permitir que se realicen todas las operaciones del circuito cuántico. Si es demasiado corto, los qubits pueden perder las propiedades de superposición y entrelazamiento, y los resultados del algoritmo van a ser incorrectos y poco confiables. Muchos algoritmos, como el de Shor (\cite{shor1994}) o el de Grover (\cite{grover1996}), dependen de dichas propiedades para lograr su ventaja sobre los métodos clásicos. Un nivel de entrelazamiento débil implica que el algoritmo no tenga ninguna ventaja cuántica u obtenga resultados incorrectos.

Las compuertas nativas son operaciones básicas del hardware cuántico. Estas se implementan directamente en la arquitectura física del procesador cuántico y son específicas del tipo de tecnología utilizada (por ejemplo, qubits superconductores o iones atrapados). Son los ''bloques de construcción'' básicos a partir de los cuales se construyen los algoritmos cuánticos más complejos (\cite{krantz2019}).

Las compuertas de Clifford (Hadamard, Pauli-X, Pauli-Z y CNOT) son operaciones cuánticas básicas que permiten manipular y entrelazar qubits. Se pueden simular eficientemente en computadoras clásicas (\cite[p. 464]{nielsen2010}) lo que implica que un dispositivo que solo utiliza compuertas Clifford y mide y prepara estados en la base computacional, no explota ninguna ventaja cuántica (\cite{Gottesman2004}). Si al grupo de Clifford se le agrega la compuerta T, se obtiene el conjunto Clifford+T el cual tiene la propiedad de ser universal. Un dispositivo cuyas compuertas nativas forman un conjunto universal tiene la capacidad de realizar, en principio, cualquier cálculo cuántico. Obviamente, las imperfecciones en los dispositivos NISQ imponen severas restricciones a la ventaja cuántica. Esto implica que en ingeniería de software, la implementación de estas compuertas debe considerarse cuidadosamente para lograr una optimización de recursos (\cite{bravyi2005}).

La conectividad de los qubits es su capacidad, determinada por la topología del hardware, para establecer interacciones entrelazantes entre sí. Decimos que dos qubits están conectados si es posible aplicar una compuerta nativa entrelazante entre ellos. En sistemas con conectividad limitada, los qubits solo pueden interactuar con sus vecinos más cercanos, lo que requiere operaciones extras (como intercambios de estado con compuertas Swap) para entrelazar qubits no adyacentes. Esto aumenta la complejidad del circuito y el tiempo de ejecución de los algoritmos y, por lo tanto, la tasa de error global. Algunas arquitecturas tienen conectividad total. Esto implica que es posible aplicar una compuerta entrelazante directamente entre cualquier par de qubits elegidos, como es el caso de los procesadores basados en trampas de iones (IonQ, AQT). Por otro lado, los dispositivos basados en qubits superconductores (IBM, Google, IQM) suelen tener conectividad limitada.

El nivel de ruido consiste en fluctuaciones aleatorias y no controladas en los parámetros físicos que interactúan con los qubits. Estas fluctuaciones pueden originarse a partir de diversas fuentes, tales como el ruido térmico (variaciones de voltaje y corriente), fluctuaciones de amplitud o fase en los osciladores que generan pulsos de control, o campos eléctricos y magnéticos fluctuantes en el entorno local del qubit, como en superficies metálicas o interfaces sustrato-metal. Estas perturbaciones generan cambios no deseados en los parámetros del qubit, lo que provoca decoherencia y reduce la fidelidad de las operaciones cuánticas. El ruido puede clasificarse en dos tipos principales: ruido sistemático, que consiste en errores predecibles y repetibles, como una rotación incorrecta de un pulso de microondas debido a una calibración imperfecta, y ruido estocástico, que consiste en fluctuaciones aleatorias e impredecibles, como las causadas por el ruido térmico o la interferencia electromagnética (\cite{krantz2019}).
 \vspace{-1em} 
\section{Gestión de recursos cuánticos hoy}
\label{section:related-work}
\vspace{-0.6em} 
Los enfoques actuales al estudio y gestión de los recursos en computadores cuánticos pueden clasificarse entre los que hacen estimación de recursos (Quantum Resource Estimation) y los que hacen evaluaciones comparativas (Benchmarking). Los primeros apuntan a determinar las necesidades de recursos de un algoritmo o sistema cuántico determinado. Los segundos buscan establecer bases de evaluación objetivas que permitan comparar tecnologías cuánticas. 
\vspace{-0.6em} 
\subsection{Estimadores de Recursos Cuánticos}
\label{qre}
\vspace{-0.6em} 
Actualmente, varios proyectos están en desarrollo dentro del área de investigación conocida como QRE (Estimador de Recursos Cuánticos). Estos proyectos se centran en estimar y optimizar los recursos necesarios principalmente para computadoras cuánticas universales tolerantes a fallos.

El estimador de recursos de Microsoft (Azure Quantum Resource Estimator) calcula el tiempo de ejecución y el número total de qubits basándose en una estimación de costo de la cantidad de compuertas necesarias para los circuitos cuánticos (\cite{AzureQRE2023}).

Google desarrolló Qualtran, que está en etapa beta. Qualtran simula y prueba algoritmos utilizando abstracciones y estructuras de datos para generar automáticamente diagramas con información y tabular los requisitos de recursos. Ofrecen una biblioteca estándar de bloques de construcción para compilaciones que minimizan costos (\cite{qualtran2024}).

Zapata AI desarrolló BenchQ como parte del programa de DARPA Quantum Benchmarking Program (\cite{darpa2023}). BenchQ ofrece herramientas para estimar los recursos de hardware necesarios en la computación cuántica resistente a errores. Cuenta con un compilador de estados gráficos, modelos de fábricas de destilación, evaluaciones de rendimiento de decodificadores, un diseño de arquitectura para trampas de iones, implementaciones de algoritmos cuánticos específicos, entre otras funcionalidades. Además, los desarrolladores aclaran que hacen uso de la misma herramienta diseñada por Microsoft Azure Quantum Resource Estimation (QRE) (\cite{benchq2025}).

Existen enfoques como el de García-Alonso (\cite{9645184}) donde proponen un modelo de Quantum Software as a Service (QSaaS) mediante una API Gateway, demostrando la necesidad de abstraer la complejidad del hardware cuántico. Mientras que herramientas como Azure Quantum Resource Estimator se centran en métricas estáticas, otros enfoques como (\cite{9645184}) adoptan un paradigma de servicios para simplificar el acceso a recursos cuánticos que incluye un backend empresarial que gestiona telemetría y ejecuciones en múltiples proveedores comerciales (IBM, D-Wave, etc.).

M. Suchara y colaboradores de la Universidad de California, Santa Bárbara, en 2013, desarrollaron una herramienta llamada QuRE. La describen como una aplicación de estimación que calcula el costo de implementaciones prácticas de circuitos cuánticos en una variedad de tecnologías cuánticas físicas, exclusivamente para la codificación en computadoras tolerantes a fallos (\cite{QuRE2013}).

Munich Quantum Toolkit (MQT) ofrece un conjunto de benchmarks para evaluar herramientas de software y automatización de diseño en computación cuántica llamado MQT Bench\label{MQT Bench}. Utilizan circuitos cuánticos pre-diseñados en cuatro niveles de abstracción, desde el nivel algorítmico hasta el nivel dependiente del hardware. Tienen análisis previos sobre cómo es conveniente la ejecución de diferentes computadoras de IBM, Rigetti, IonQ, OQC (Oxford Quantum Circuits) y Quantinuum, teniendo en cuenta sus compuertas nativas.  Dan soporte para acceder mediante una interfaz web y un paquete de Python. Hacen uso del compilador de Qiskit y/o TKET y sus optimizadores por defecto. Tienen cinco conjuntos de compuertas nativas y siete dispositivos cuánticos con capacidades que van de 8 a 127 cúbits. Su objetivo es mejorar la comparabilidad, reproducibilidad y transparencia en la evaluación de herramientas de software cuántico (\cite{MQTBenchBenchmarking2023}).

\sloppy
Otro trabajo, realizado por Saadatmand (\cite{FTQCs2024}), titulado ''Fault-tolerant resource estimation using graph-state compilation on a modular superconducting architecture'', se enfoca en la estimación de recursos necesarios para la computación cuántica tolerante a fallos (FTQC). Los autores emplean una metodología basada en la compilación de estados gráficos (graph-state compilation) para evaluar los requisitos de recursos en sistemas FTQC. Analizan en arquitectura superconductora cómo la conectividad entre módulos, la latencia y otros factores impactan en los recursos totales necesarios para implementar algoritmos cuánticos.

En un trabajo realizado para DARPA y la NASA, Mozgunov, Marshall y Anand (\cite{DarpaNasa2024}) describen dos aplicaciones en las que su enfoque preferido es la simulación cuántica de sistemas abiertos. Proponen optimizaciones algorítmicas con parámetros específicos para las estimaciones de recursos aprovechando la invarianza traslacional, propiedad de un sistema o una función matemática de permanecer inalterada cuando se desplaza o traslada en el espacio que simplifica cálculos aprovechando estructuras que se repiten periódicamente, además del paralelismo en la aplicación de la compuerta T. Se  implementan únicamente en FTQC. Las estimaciones de recursos para estos sistemas dependen centralmente de la huella del estado T, es decir, de los recursos necesarios para generar y mantener estados T.

\vspace{-0.9em}

\subsection{Evaluación comparativa}
\label{benchmarking}

\vspace{-0.7em}  
En los proyectos descritos en la sección anterior, el enfoque principal para la estimación de recursos depende del algoritmo, su funcionalidad y propósito, o de los datos previos que tenemos sobre la computadora cuántica proporcionada por el proveedor de servicios. Para evaluar el hardware, también existen muchos proyectos de benchmarking basados en diferentes factores (\cite{BenchmarkingQuantumComputers2025}). Esto también influye en los resultados de las pruebas, anuncios y conclusiones, tanto propias como en las de los competidores.

\vspace{-1em} 

\begin{table}[H]
\centering
\caption{Comparación de herramientas para estimación y benchmarking en computación cuántica}
\begin{tabular}{|p{1.5cm}|p{1.7cm}|p{2cm}|p{0.9cm}|p{1.1cm}|p{2.5cm}|p{2cm}|}
\hline
\textbf{Proyecto} & \textbf{Tipo} & \textbf{Métricas} & \textbf{NISQ} & \textbf{Open \newline Source} & \textbf{Tecnología} & \textbf{Ref} \\
\hline
Azure \newline Quantum QRE & Estimación & Qubits, \newline compuertas T, tiempo de ejecución & No & No & FTQC \newline (super\newline conductores) & \cite{AzureQRE2023} \\
\hline
Qualtran (Google) & Estimación & Profundidad de circuitos, qubits físicos y lógicos & Parcial & No & FTQC/NISQ (genérico) & \cite{qualtran2024} \\
\hline
BenchQ & Bench \newline marking & Fidelidad, conectividad, recursos para FTQC & Sí & Sí & FTQC \newline (trampa \newline de iones) & \cite{benchq2025} \\
\hline
\end{tabular}
\label{tab:herramientas-qre}
\end{table}

\begin{table}[H]
\centering
\begin{tabular}{|p{1.5cm}|p{1.7cm}|p{2cm}|p{0.9cm}|p{1.1cm}|p{2.5cm}|p{2cm}|}
\hline
\textbf{Proyecto} & \textbf{Tipo} & \textbf{Métricas} & \textbf{NISQ} & \textbf{Open \newline Source} & \textbf{Tecnología} & \textbf{Ref} \\
\hline

QuRE & Estimación & Costo de circuitos en FTQC, compuertas nativas & No & No & FTQC \newline(diversas\newline tecnologías) & \cite{QuRE2013} \\
\hline
MQT \newline Bench & Bench\newline marking & Compuertas nativas, ruido, \newline compatibilidad & Sí & Sí & NISQ \newline (multi-plataforma) & \cite{MQTBenchBenchmarking2023} \\
\hline
FTQC \newline graph-state compilation & Estimación & Recursos para FTQC, modularidad, conectividad & No & No & FTQC \newline (super \newline conductores\newline modulares) & \cite{FTQCs2024} \\
\hline
DARPA \newline NASA & Estimación & Huella de estados T, optimización algorítmica & No & No & FTQC \newline (simulación abierta) & \cite{darpa2023} \\
\hline
Quantum Path\textsuperscript{\textregistered} & Plataforma & Agnosticismo, integración híbrida, \newline telemetría & Sí & No & Plataformas \newline comerciales como IBM, D-Wave, Rigetti, AWS Braket, etc. & \cite{9645184} \\
\hline
\end{tabular}
\label{tab:herramientas-qre2}
\end{table} 
\vspace{-1em} 
En 2018, el equipo de Google diseñó un método de benchmarking para la supremacía cuántica  (\cite{boixo2018},\cite{bravyi2005}). Este método se fundamenta en estudios teóricos sobre la dificultad de muestreo a partir de circuitos cuánticos aleatorios, la dureza computacional de su simulación clásica y sus implicaciones en la demostración de la supremacía cuántica, estableciendo conexiones con la complejidad computacional y las separaciones entre modelos cuánticos y clásicos.(\cite{aaronson2017})

La implementación experimental del muestreo sobre circuitos cuánticos aleatorios llevó a Google a anunciar que el procesador Sycamore tardaba aproximadamente 200 segundos en muestrear un millón de veces una instancia de un circuito cuántico. Indicaban que la misma tarea para una supercomputadora clásica de última generación tardaría aproximadamente 10,000 años y que esta supremacía cuántica también anunciaba la era de las tecnologías cuánticas de escala intermedia ruidosa (NISQ). Además, afirmaron que con ese benchmarking se aseguraba una aplicación inmediata en la generación de números aleatorios certificables (\cite{sycamore2019}). Poco después, IBM (\cite{pednault2019}) propuso una simulación mucho más eficiente con hardware clásico para realizar la misma ejecución en 2.5 días y no en 10,000 años.
Esta experiencia demuestra que hay que ser cuidadosos en la selección de las pruebas y tests a realizar para sacar conclusiones correctas.

No todo Benchmarking es útil en todos los casos. En la referencia \cite{BenchmarkingQuantumComputers2025} se describen múltiples métodos disponibles, detallando cuáles son más adecuados para la era NISQ, otros para etapa fault-tolerant, de bajo nivel, alto nivel y cómo esto va a tener que ir cambiando necesariamente según el progreso tecnológico. También describen seis propiedades fundamentales para lograr un buen benchmark:

\begin{enumerate}
    \item Deben estar bien fundamentados, justificando las métricas de rendimiento.

    \item Un procedimiento tiene que ser claro y sin ambigüedades. Cualquier paso no especificado en ese procedimiento debe ser un parámetro configurable intencionadamente. 

    \item En su implementación no debería ser posible manipular los parámetros configurables para obtener resultados engañosos. 

    \item Los resultados no deberían corromperse cuando la computadora cuántica evaluada experimente errores pequeños o no previstos. 

    \item Tienen que ser eficientes y no utilizar una gran cantidad de recursos. 

    \item Por último, el benchmark debe ser independiente de la tecnología; solamente debería ser particular si lo que se va a medir es especial de esa arquitectura.
\end{enumerate}

\vspace{-0.6em} 

\section{Desafíos en la gestión de recursos cuánticos en la era NISQ}
\label{s:Recursos_en_NISQ}

\vspace{-0.6em} 

Desde el punto de vista de la Ingeniería de Software, determinar los recursos disponibles en los prototipos de computadoras cuánticas, independientemente de la plataforma y de manera confiable para poder hacer una gestión correcta, sigue siendo un desafío por resolver.

Las técnicas de pruebas referidas en \ref{benchmarking} y librerías o herramientas detalladas en \ref{qre}, pueden ser útiles en diferentes casos. Aunque, además de que en su mayoría solo funcionan únicamente con CCUTEs, la aplicación de los mismos debe hacerse con anterioridad a la ejecución, tanto para la verificación de los requerimientos, como para la ejecución. Es una visión estática del uso de las computadoras cuánticas que no considera la particularidad de cómo funcionan ni la etapa de desarrollo tecnológico de utilidad en la que nos encontramos. En este último punto podemos hacer la salvedad con MQT Bench \ref{MQT Bench}, que se basa en datos previamente obtenidos, pero no actualizados al momento de ejecutar el algoritmo. La conectividad de los qubits puede ser la misma si no cambia la topología del hardware, pero su nivel de entrelazamiento, su tiempo de coherencia o fidelidad pueden cambiar de un momento a otro.
    
Para un uso eficiente que tome en consideración los recursos utilizables y que se adapte a las necesidades del usuario, será necesaria una capa como parte del mismo software a ejecutar. En este trabajo remarcamos la necesidad de desarrollar una capa que tenga presente el flujo de trabajo del mismo algoritmo, verificando que los recursos estén disponibles de manera dinámica, independientemente del hardware o del servicio cuántico disponible y que sean efectivos para el uso proyectado.

Existen recursos numerables e identificables de manera sencilla en una instancia previa a la ejecución, como cantidad de qubits necesarios, fidelidad, tiempo de coherencia, compuertas disponibles, topología según el hardware. (\cite{Saraiva2021}. La información sobre estas características es usualmente proporcionada por el proveedor del servicio (aunque la frecuencia con la cual está disponible de forma actualizada puede variar de proveedor a proveedor).

Sin embargo, lo esperable para obtener ventaja cuántica es la capacidad de generar estados con los recursos físicos necesarios, apelando a un conjunto de compuertas nativas universales. Las capacidades que pueda llegar a tener un dispositivo dado para cumplir con ese objetivo pueden variar con el tiempo. La sensibilidad al ruido de los qubits puede perjudicar la ejecución y fiabilidad en los resultados del circuito/algoritmo. En particular, los estados logrados a partir de la utilización del conjunto de compuertas universales de cada dispositivo tienen que ser capaces de generar estados mágicos o ''Non-stabilizerness'', los cuales son los que permiten que los algoritmos cuánticos superen las capacidades clásicas. En un trabajo del 2025 (\cite{macedo2025}) hacen uso de las desigualdades de Bell (\cite{brunnerBellNonlocality2014}) especialmente diseñadas para que puedan actuar como testigos de estos estados mágicos.
En otros trabajos de investigación previos (\cite{holik2024,arango2024}) \label{investigaciones} se analizan correlaciones cuánticas mediante diferentes medidas de entrelazamiento, comparando conjuntos de puertas universales a partir de estados generados por circuitos aleatorios cuánticos (QRC). Por otro lado, es importante remarcar la necesidad de desarrollar software de testeo que no implique exponer el algoritmo a usar. Las herramientas presentadas en (\cite{holik2024,arango2024}) son un paso importante para cumplir con ese objetivo, dado que la caracterización de correlaciones usando circuitos cuánticos aleatorios permite tener una evaluación global de un conjunto de qubits dado.

\section{Conclusiones y trabajo futuro}\label{s:Conclusiones}

Concluimos que es necesaria una capa que evalúe los recursos disponibles para resolver de manera dinámica si es conveniente proceder a la ejecución, brindar una mayor confianza en los resultados obtenidos y garantizar la funcionalidad eficiente de esta tecnología. 
\begin{figure}[!ht]
    \centering
    \includegraphics[width=0.6\textwidth]{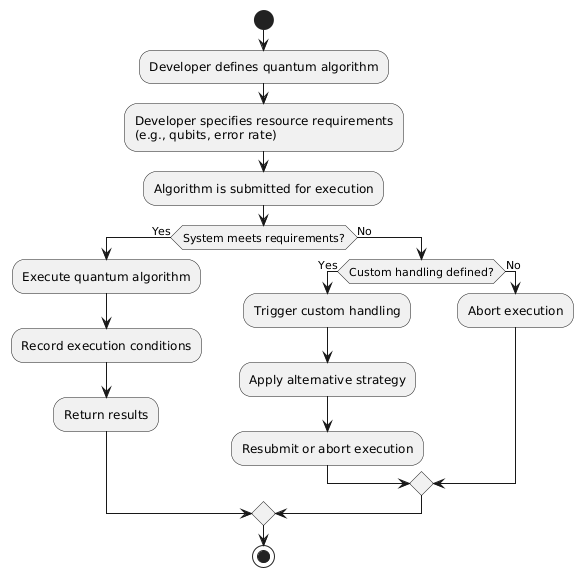}
    \caption{Diagrama de flujo de una capa  de integración}
    \label{fig:Workflow}
\end{figure}
Haciendo uso de las investigaciones mencionadas en \ref{investigaciones}, es posible evaluar si los dispositivos disponen de los recursos requeridos por el usuario final con una capa dinámica (Figura 1) extendiendo mediante evaluación en tiempo real. Esta información puede ser útil a la hora de decidir entre diferentes prototipos de computadoras cuánticas.

A futuro, planificamos implementar esta propuesta como una capa de integración, la cual, en términos generales, presentaría las siguientes características:

\begin{enumerate}
    \item Ofrecer una librería de funciones para consultar recursos disponibles, expresar requisitos y/o restricciones de código abierto.
    \item Ser una librería abstracta, permitiendo implementaciones específicas para distintas computadoras cuánticos.
    \item Ser extensible, de modo que se pueden agregar nuevas abstracciones de recursos a medida que evoluciones las plataformas.
    \item Incluir un lenguaje propio para definir el manejo de las excepciones (por ejemplo, cuando no se cumple con determinado requerimiento).
\end{enumerate}

{\scriptsize
\begin{table}[H]
\centering
\caption{Comparación entre la propuesta actual y herramientas existentes}
\begin{tabular}{|p{2.5cm}|p{5.5cm}|p{5.3cm}|}
\hline
\textbf{Criterio} & \textbf{Propuesta} & \textbf{Herramientas Existentes} \\
\hline
Enfoque \newline Dinámico & Evaluación en tiempo real durante la ejecución (ruido, qubits disponibles, etc.). Monitorización de métricas NISQ. & Análisis estático previo. Datos históricos/fijos (ej: MQT Bench). \\
\hline
Soporte NISQ & Decisión adaptativa basada en estado actual del hardware (ej: tiempo de coherencia residual). & Mayoría orientada a FTQC. Soporte NISQ limitado sin gestión dinámica). \\
\hline
Agnosticismo & Librería abstracta adaptable a cualquier hardware. Extensible (métricas personalizables). & QuantumPath\textsuperscript{\textregistered} agnóstico pero cerrado. No extensibles). \\
\hline
Integración & Middleware independiente (Fig. \ref{fig:Workflow}). Manejo de excepciones. Código abierto. & Soluciones monolíticas. Sin retroalimentación proactiva (benchmarks solo reportan métricas). \\
\hline
\end{tabular}
\label{tab:comparativa-propuesta}
\end{table}
}
\clearpage
\printbibliography 

\end{document}